\documentstyle[aps,12pt]{revtex}

\begin{document}
\title{Relativistic effects in the electronic structure\\
for the 3d paramagnetic ions$^{***}$ }
\author{R.J. Radwa\'{n}ski}
\address{Center for Solid State Physics, \'{s}w. Filip 5, 31-150 Krak\'{o}w,\\
Inst. of Physics, Pedagogical University, 30-084 Krak\'{o}w, Poland.}
\author{Z. Ropka}
\address{Center for Solid State Physics, \'{s}w. Filip 5, 31-150 Krak\'{o}w\\
email: sfradwan@cyf-kr.edu.pl}
\maketitle

\begin{abstract}
It has been shown that relativistic spin-orbit effects have enormous
influence on the electronic structure of the correlated electron systems 3d$%
^n$. The s-o coupling produces the fine electronic structure with a large
number of low-energy, 
\mbox{$<$}
10 meV, localized states, These states affect enormously electronic and
magnetic properties of 3d-ion compounds at low- and room-temperature
regions. It is inferred that the relativistic s-o effects and the multiplet
structure on individual atoms are indispensable for the proper evaluation of
the electronic structure of paramagnetic 3d ions.
\end{abstract}

\date{(30 May 1997)}

The 3d-paramagnetic-ion compounds are the subject of continuous interest$%
^{1-17}$ despite of more than 60 years of intensive theoretical and
experimental studies. The relativistic spin-orbit (s-o) coupling is usually
ignored in calculation of the electronic structure of the 3d ions in solids
basing on a general consensus that the s-o coupling is for 3d ions
relatively weak. We have argued$^{18}$~that it is just opposite - the weak
s-o coupling causes dramatic change of the electronic structure by producing
the fine electronic structure with low-energy excitations even so small as 1
meV (=11.6 K = 8.0 cm$^{-1}$).

The aim of this Letter is to present the influence of the spin-orbit
coupling on the localized states of the highly-correlated d$^n$ electronic
system produced by crystal-field interactions of the cubic octahedral
symmetry. The present approach is an extension of an approach known as an
intermediate crystal-field approach$^{17}$. One can be surprised, but the
influence of the spin-orbit coupling has not been systematically studied$%
^{19}$ despite of a quite simple form of the s-o Hamiltonian $H_{s-o}$ = $%
\lambda L\cdot S$. It turns out that the spin-orbit coupling has to be taken
into account for any meaningful analysis of electronic and magnetic
properties of 3d-ion compounds due to the presence of the fine electronic
structure.

The modern approach to a 3d-ion compound is based on an idea of Mott that it
is strong electronic correlations that make electrons in the incomplete 3d
shell to stay rather localized than itinerant (Mott insulators). The
physical situation for the 3d$^n$ system of a 3d-transition-metal ion is
here taken to be accounted for by considering the single-ion-like
Hamiltonian containing the electron-electron d-d interactions $H_{el-el}$,
the crystal-field $H_{CF}$, spin-orbit $H_{S-O}$, and Zeeman $H_Z$~
interactions$^{20}$

\begin{center}
$H_d=H_{el-el}+H_{CF}+H_{s-o}+H_Z$ (1).
\end{center}

The electron-electron and spin-orbit interactions are intra-atomic
interactions, whereas crystal-field and Zeeman interactions account for
interactions of the unfilled 3d shell with the charge and spin surroundings.
These interactions are written in the decreasing- strength succession.$^{18}$%
~

In a zero-order approximation the electron-electron correlations are
accounted for by phenomenological Hund's rules (the maximal value of L
provided the maximal value of the spin S is realized). They yield for the 3d$%
^n$ electron configuration the ground term~$\left| \text{LS}\right\rangle $
that is (2L+1)$\cdot $(2S+1) degenerated, see Table 1. Strong intra-atomic
Hund's-rule correlations allows one to work in the~$\left| \text{LSL}_Z\text{%
S}_Z\right\rangle $ space. In this space the effect of the crystal-field, of
the spin-orbit coupling and of the (internal/external) magnetic field is
accounted for by considering the single-ion Hamiltonian of the 3d$^n$ system
of the 3d-transition-metal ion of the form$^{14}$:

\begin{center}
$H_d=$B$_4(O_4^0+$5$O_4^4)+\lambda L\cdot S+\mu _B(L+$g$_eS)\cdot B_{ext}$
(2)$.$
\end{center}

The first term is the cubic CEF Hamiltonian with the Stevens operators $%
O_n^m $ that depend on the orbital quantum numbers L, L$_z$. The second term
accounts for the spin-orbit interactions. The last term accounts for the
influence of the magnetic field, the externally applied in the present case.
g$_s$ value is taken as 2.0023. The computations of the many-electron states
of the 3d$^n$ system have been performed by consideration of the Hamiltonian
(3) in the $\left| \text{LSL}_Z\text{S}_Z\right\rangle $ base$^{21}$. As a
result of the diagonalization one obtains the energies of the (2L+1)$\cdot $%
(2S+1) states and the eigenvectors containing information e.g. about the
magnetic properties. These magnetic characteristics are computationally
revealed under the action of the external magnetic field B$_{ext}$.

Fig. 1 presents the calculated general overview of the CEF and spin-orbit
effect on the Hund's rule term for the d$^n$ systems. In figures b the
splitting of the ground term by the octahedral cubic CEF interactions is
shown for different number of n. Figures c show the splitting due to the
spin-orbit interactions. The parameters used are collected in Table 1. They
have been chosen in order to have the CEF splitting of about 2 eV in
agreement with the quite frequent experimental observations of the d-d
excitations in the optically visible energy range.$^{1-9}$

Figure 1 is full of information. One can see the similarities, but also the
differences, between the CEF effect and s-o interactions on, for instance,
the $^2$D and $^5$D terms for the d$^1$ ($\lambda $ 
\mbox{$>$}
0) and d$^6$ ($\lambda $ 
\mbox{$<$}
0) terms. The particle-hole symmetry can be studied for the d$^n$/d$^{10-n}$
systems. We mention here only a few most important points.

\begin{enumerate}
\item  For 3d ions there are only D and F terms as L can be only 2 and 3.
The spin degeneracy depends on the number of electrons involved. The total
degeneracy as large as 15 can be realized.

\item  The pairs d$^1$/d$^6$, d$^2$/d$^7$, d$^3$/d$^8$ and d$^4$/d$^9$ of
systems have the same orbital ground state. The spin-orbit effect is,
however, entirely different due to different values of S, the reversal of
the s-o constant $\lambda $ and the transformation of non-Kramers ions into
Kramers ions (even~$\Longleftrightarrow $ odd number of d electrons).

\item  The adoption or the loss of the one electron draws the full
reconstruction of the fine electronic structure.

\item  The spin-orbit coupling removes largely the degeneracy in all cases
apart of the spin-only d$^5$~ system, see Table 1 as well as Fig.1. In case
of the d$^4$ system it yields a singlet ground state$^{18}$ , but there are
10 closely lying levels.

\item  The s-o coupling produces a fine electronic structure with
excitations below 10 meV. These low-energy excitations can be detected by
specific-heat experiments, for instance.

\item  There is splitting of the $^5$E$_g$ cubic state$^{22}$, the fact that
is of particular importance for the d$^4$ system as it becomes the ground
state.

\item  There is also the extra second-order splitting of the T$_{2g}$
originated states in comparison to the perturbation method.

\item  The ground-state magnetic moment, shown in Fig.1, substantially
differs from the spin-only value. A very small moment has been revealed for
the d$^1$, d$^2$~ and d$^4$ systems. In all cases it differs much from an
integer value.

\item  The states shown are {\bf many-electron states} of the whole 3d$^n$
system. At {\bf 0 K only the lowest state is occupied. Higher excited states
become populated with increasing temperature.}

\item  The population of higher states manifests in temperature variation of
electronic and magnetic properties like the specific heat and the magnetic
susceptibility. Detailed calculations of $\chi $(T) have been presented
already for the d$^6$ system.
\end{enumerate}

From physical point of view the most important seems to be the point 9 and
3. It shows the fundamental difference with the very often recalled
one-electron picture with subsequent occupation of the t$_{2g}$ and e$_g$%
~states. In the many-electron picture at 0 K only the lowest state is
occupied, i.e. one can say that all n electrons as the whole d$^n$ system is
put on the lowest level. In the one-electron picture electrons are put
subsequently$^{6-13}$ on the 10 CEF levels without the reconstruction of the
electronic structure in contrast to the present results, see point 3. The
evaluation of the fine electronic structure, with the given energy and
magnetic characteristics, allows for the calculation of temperature
dependence of many physical properties like specific heat or the
paramagnetic susceptibility similarly to the desprition used for the
rare-earth compounds.$^{23}$

According to us this many-electron picture is widely confirmed though the
one-electron picture is often recalled in discussion of 3d-ion compounds$%
^{6-13}$. Here we mention EPR experiments, where -provided the experiment is
performed at, at least, helium temperature - the ground-level properties can
be revealed. For the Ni$^{2+}$ ion, where the d$^8$ system is realized, the
spectroscopic g-value of 2.15-2.35 (AB p. 449; it corresponds to the same
value of the magnetic moment in $\mu _{\text{B}}$~as S=1) is commonly found
for the octahedral interstice. It is close to the value of 2.15 shown in
Fig. 1 for the d$^8$ system. Larger values can be obtained by slightly
different values for B$_4$, $\lambda $ and by a lattice distortion. Also a
value of 3.52 calculated for the 3d$^6$ system seems to have been seen in Fe$%
^{2+}$ impurities in MgO (the octahedral interstice), where value of 3.43
has been reported (AB p. 444). In ref. 14 the g-factor experimentally
observed for the Cu$^{2+}$ ions has been calculated reaching very good
agreement.

The present calculations are the exact calculations. Up to now the s-o
coupling, if discussed, has been taken into account by the perturbation
method, with the first-order effect in $\lambda $.

{\bf \ In conclusion}, large influence of the spin-orbit coupling on the
localized states of the d$^n$ system has been revealed. The present approach
treats n d electrons as the {\bf whole highly-correlated d}$^n${\bf \
system. } The s-o coupling, in combination with crystal-field interactions,
produces the fine electronic structure with a large number of low-energy, 
\mbox{$<$}
10 meV, localized states. These states affect enormously electronic and
magnetic properties of 3d-ion compounds at low- and room- temperature
regions. The present calculations prove that the weaker s-o coupling the
smaller energy region for the fine low-energy electronic structure and the
lower temperatures with the occurrence of anomalies in physical properties.
On basis of these studies it is inferred that the relativistic s-o effect
and the multiplet structure on individual atoms are indispensable for the
proper evaluation of the electronic structure of paramagnetic 3d ions. It is
believed that the found fine local electronic structure is essential for the
low-and room-temperature properties of 3d-ion compounds and have to be
incorporated in modern band-structure calculations.

Note added during the referee process. The presently-published calculations
of properties of 3d-ion compounds do not take the s-o coupling into account,
e.g. Phys.Rev.B 54 (1996) 5309 on p. 5315 right column, line 12 bottom;
Phys.Rev.B 53 (1996) 7158 on p.7164, left column, line 16 bottom. Those
authors say that in their approaches the s-o effect is small. Our results
are completely different. Arguments of the referee, that ''the results
presented do not appear to be useful, in the sense that they might explain
phenomena in particular 3d systems that have not been understood'' and that
''the effects predicted occur at small energy and no case has been made by
the author that they influence importantly the physical properties of a know
system'' are incorrect. These results e.g. explain non-integer values of the
3d magnetic moment, enable evaluation of the orbital moment, offer
explanation for the formation of the non-magnetic state of LaCoO$_3$ (the
zero-value state for the d$^6$ system in Fig.1) and enable calculation of
temperature dependence of the paramagnetic susceptibility that turns out to
be very anomalous. All of these outcomes are in agreement with experimental
results. The energy splitting of e.g. the $^3$T$_{1g}$ state for the d$^2$
system amounts to 3$\lambda .$ It means that in the energy window of 450 K
(40 meV) 9 levels are expected. Thus we expect that a d$^2$ system will show
anomalies at low and room-temperature regions. These temperature regions are
of great importance for the solid-state physics.

$^{***}$This paper has been submitted to Phys.Rev.Lett. (LE6925) 30.05.1997
but up to July 1999 did not get appreciation of the referees and the Editor
of Phys. Rev.Lett. despite of numerous elaborated answers to referee reports
pointing out that the paper presents another scientific point of view then
the referees about the role of the intra-atomic spin-orbit coupling. Part of
this paper has been presented at VII Polish Conf. on High-Tc
superconductivity in Miedzyzdroje 1-3.09.1997 (Poland, organized by
H.Szymczak et al.) but also did not get appreciation of the Scientific
Committee chaired by H.Szymczak. This paper has been distributed as the
Report of Center for Solid State Physics CSSP-4/97.

{\bf Table 1.} Spin S and orbital L quantum numbers of the ground term for
the highly-correlated 3d$^n$ electronic systems with the total degeneracy in
the LS space. The ground term and the degeneracy in the octahedral cubic
crystal field with the respective values of the cubic CEF\ parameter B$_4$.
Next, the spin-orbit coupling parameter $\lambda $ and the degeneracy
resultant from the spin-orbit coupling in the presence of the octahedral
cubic CEF\ interactions. These values of B$_4$ and $\lambda $ have been used
for calculations of Fig.1.

\begin{tabular}{cccccc}
n & S & L & 
\begin{tabular}{c}
Free ion \\ 
\begin{tabular}{cc}
Ground Term & degeneracy
\end{tabular}
\end{tabular}
& 
\begin{tabular}{c}
octa cubic CEF \\ 
\begin{tabular}{ccc}
Ground Subterm & degeneracy & B$_4$ (K)
\end{tabular}
\end{tabular}
& 
\begin{tabular}{c}
s-o coupling \\ 
\begin{tabular}{cc}
$\lambda $ (K) & degeneracy
\end{tabular}
\end{tabular}
\\ 
d$^1$ & $\frac 12$ & 2 & 
\begin{tabular}{cc}
$^2$D & 10
\end{tabular}
& 
\begin{tabular}{ccc}
$^2$T$_{2g}$ & 6 & +200
\end{tabular}
& 
\begin{tabular}{cc}
+220 & 4
\end{tabular}
\\ 
d$^2$ & 1 & 3 & 
\begin{tabular}{cc}
$^3$F & 21
\end{tabular}
& 
\begin{tabular}{ccc}
$^3$T$_{1g}$ & 9 & -40
\end{tabular}
& 
\begin{tabular}{cc}
+150 & 2
\end{tabular}
\\ 
d$^3$ & $\frac 32$ & 3 & 
\begin{tabular}{cc}
$^4$F & 28
\end{tabular}
& 
\begin{tabular}{ccc}
$^4$A$_{2g}$ & 4 & +40
\end{tabular}
& 
\begin{tabular}{cc}
+125 & 4
\end{tabular}
\\ 
d$^4$ & 2 & 2 & 
\begin{tabular}{cc}
$^5$D & 25
\end{tabular}
& 
\begin{tabular}{ccc}
$^5$E$_g$ & 10 & -200
\end{tabular}
& 
\begin{tabular}{cc}
+120 & 1
\end{tabular}
\\ 
d$^5$ & $\frac 52$ & 0 & 
\begin{tabular}{cc}
$^6$S & 6
\end{tabular}
& 
\begin{tabular}{ccc}
- & 6 & -
\end{tabular}
& 
\begin{tabular}{cc}
- & 6
\end{tabular}
\\ 
d$^6$ & 2 & 2 & 
\begin{tabular}{cc}
$^5$D & 25
\end{tabular}
& 
\begin{tabular}{ccc}
$^5$T$_{2g}$ & 15 & +200
\end{tabular}
& 
\begin{tabular}{cc}
-140 & 3
\end{tabular}
\\ 
d$^7$ & $\frac 32$ & 3 & 
\begin{tabular}{cc}
$^4$F & 28
\end{tabular}
& 
\begin{tabular}{ccc}
$^4$T$_{1g}$ & 12 & -40
\end{tabular}
& 
\begin{tabular}{cc}
-260 & 2
\end{tabular}
\\ 
d$^8$ & 1 & 3 & 
\begin{tabular}{cc}
$^3$F & 21
\end{tabular}
& 
\begin{tabular}{ccc}
$^3$A$_{2g}$ & 3 & +40
\end{tabular}
& 
\begin{tabular}{cc}
-480 & 3
\end{tabular}
\\ 
d$^9$ & $\frac 12$ & 2 & 
\begin{tabular}{cc}
$^2$D & 10
\end{tabular}
& 
\begin{tabular}{ccc}
$^2$E$_g$ & 4 & -200
\end{tabular}
& 
\begin{tabular}{cc}
-1200 & 4
\end{tabular}
\\ 
d$^{10}$ & 0 & 0 & 
\begin{tabular}{cc}
$^1$S & 1
\end{tabular}
& 
\begin{tabular}{ccc}
- & 1 & -
\end{tabular}
& 
\begin{tabular}{cc}
- & 1
\end{tabular}
\end{tabular}

\begin{center}
{\bf Figure captions }
\end{center}

{\bf Fig. 1. }The general overview of the splitting of the Hund's rules
ground term of the 3d$^n$ electronic systems by octahedral cubic
crystal-field interactions (b) and the spin-orbit coupling (c). Levels in
(c) are labeled with degeneracies in the LS space whereas in (b) the
degeneracy is shown by the orbital spin degeneracy multiplication. The
spin-orbit splittings are drawn not in the energy scale that is relevant to
CEF levels shown in figures b. On the lowest localized level the magnetic
moment (in $\mu _{\text{B}}$~) is written. The shown states are many
electron states of the whole system d$^n$. At zero temperature only the
lowest state is occupied. The higher states become populated with the
increasing temperature.


\begin{references}
\bibitem{19}  In ref. 11 authors have tried to consider the s-o coupling for
the d$^6$ system, but in a very invalid approximate way within an effective
Hamiltonian with L'=1 instead of L=2.
\bibitem{21}  the computer program is available on the written request to
the author.
\bibitem{22}  The splitting of the $^5$E$_g$ state is a surprising finding -
the author has in hands a statement of a referee of one of the most
prestigious physical journal, that $^5$E$_g$ state is not affected by the
s-o coupling. It is true for the $^2$E$_g$ state but not for the $^5$E$_g$
state.
\end{references}
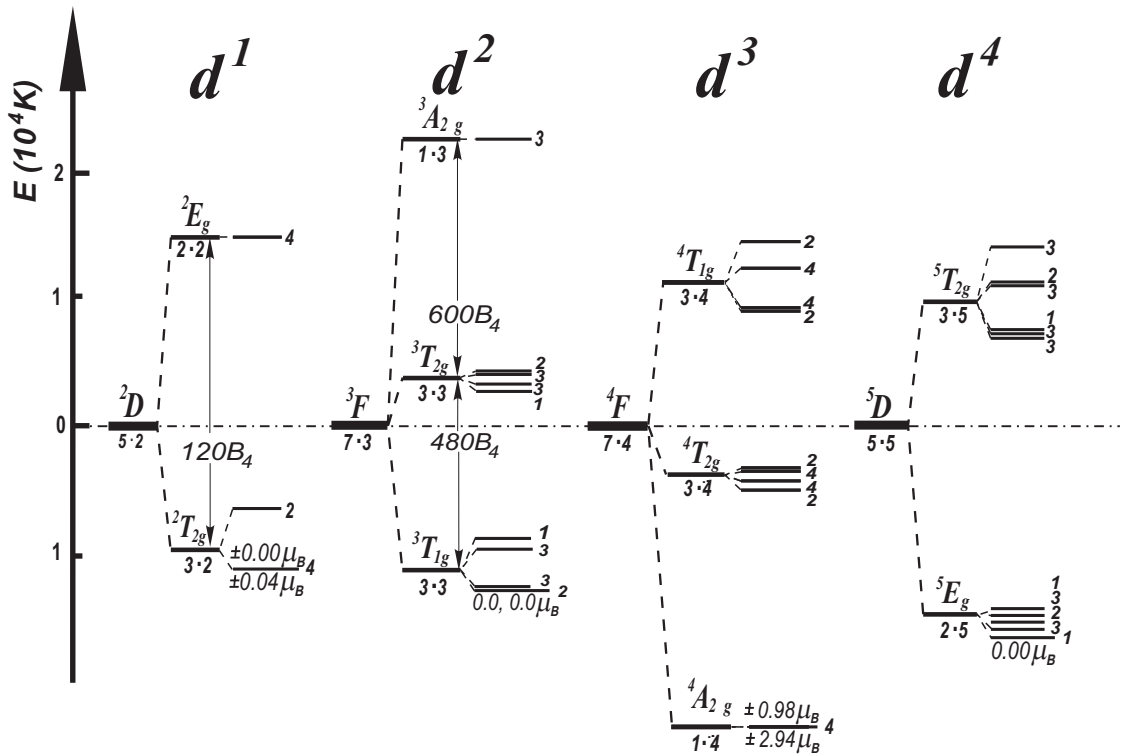
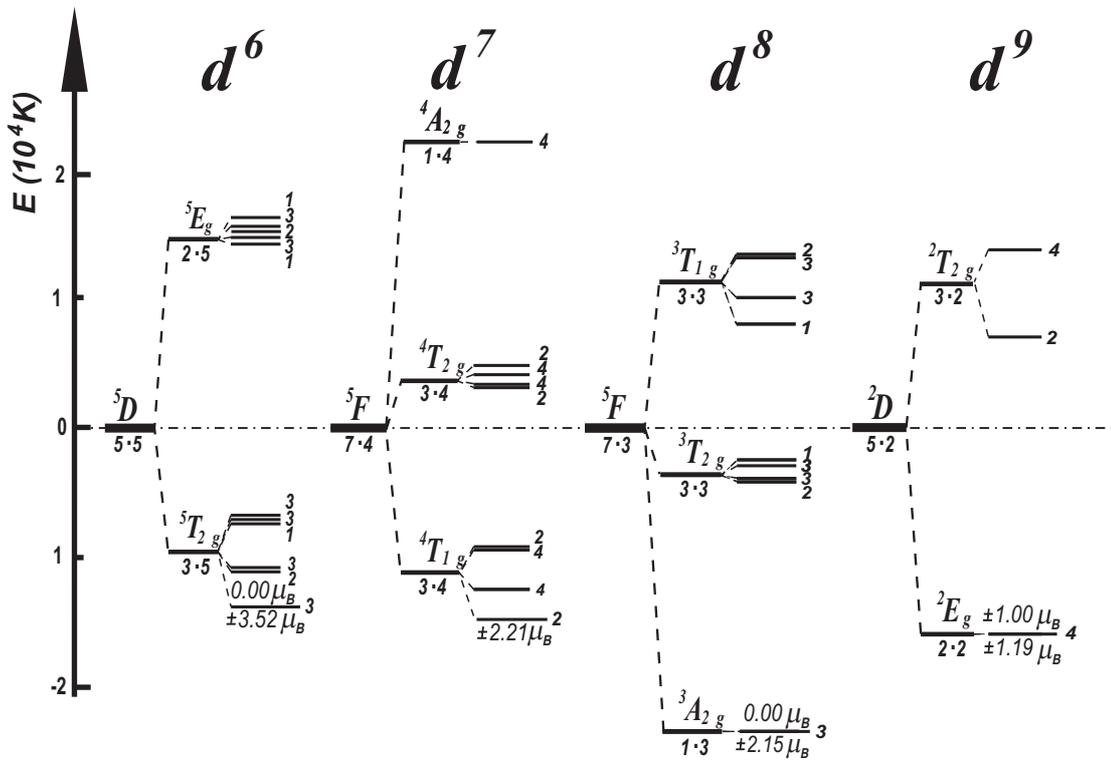
\end{document}